\documentclass{amsart}

\usepackage{amsmath,amsfonts,amssymb,amsthm,mathrsfs,graphicx,cite}
\usepackage{tikz-cd}

\begin{document}

\newtheorem{theorem}{Theorem}[section]
\newtheorem{corollary}[theorem]{Corollary}
\newtheorem{lemma}[theorem]{Lemma}
\newtheorem{proposition}[theorem]{Proposition}
\newtheorem{example}[theorem]{Example}
\theoremstyle{remark}
\newtheorem{remark}{Remark}[section]

\newcommand{\ii}{\mathrm{i}}

\title{Calogero--Moser--Sutherland systems}

\author{Martin Halln\"as}
\email{hallnas@chalmers.se}
\address{Department of Mathematical Sciences, Chalmers University of Technology and the University of Gothenburg, SE-412 96 Gothenburg, Sweden}

\date{\today}

\begin{abstract}
We discuss integrable many-body systems in one dimension of Calogero--Moser--Sutherland type, both classical and quantum as well as nonrelativistic and relativistic. In particular, we consider fundamental properties such as integrability, the existence of explicit solutions as well as action-angle and bispectral dualities that relate different such systems. We also briefly discuss the early history of the subject and indicate connections with other integrable systems.
\end{abstract}


\maketitle

\section{Introduction}

Calogero--Moser--Sutherland (CMS) systems constitute a class of integrable one-dimensional many-body systems that can be studied in great detail using exact methods, both at the classical and at the quantum level. In their original form, they describe an arbitrary number of point-particles moving on either a line or on a circle and interacting pairwise through a potential proportional to the inverse square of the distance between the particles.

The first example is due to Calogero (1971), who considered the quantum system of $N$ identical particles on the line with pair potential of the form $\omega^2 (x_i-x_j)^2+g^2/(x_i-x_j)^2$. He made the striking discoveries that the energy spectrum can be computed explicitly and that, in the $\omega=0$ case, the system exhibits particularly simple soliton scattering (conservation of momenta and factorisation of the $S$-matrix). Shortly thereafter, Sutherland (1971) initiated a study of the quantum problem on the circle, characterised by a pair potential $g^2/4\sin^2((x_i-x_j)/2)$, obtaining exact and remarkably explicit results, including a formula for the energy spectrum and an algorithm for the construction of the corresponding eigenfunctions.

On the classical level, Moser (1975) proved integrability of both systems by providing Lax representations and solved the scattering problem for the former system (with $\omega=0$), thus generalising earlier results of Marchioro (1970) from $N=3$ to all particle numbers $N>3$.

Not long thereafter, Olshanetsky and Perelomov (1977, 1983) advanced the subject substantially when they tied in the classical CMS systems with geodesic flows and the quantum systems with harmonic analysis on symmetric spaces, leading them to introduce integrable generalisations of the systems studied by Calogero, Moser and Sutherland in which the structure of the particle interactions are encoded in an arbitrary root system. In this setting, the original systems correspond to root systems of type $A_{N-1}$.

Another fundamental and related class of integrable $N$-particle systems were introduced by Ruijsenaars and Schneider (1986) in the classical- and Ruijsenaars (1987) in the quantum case. These systems can be viewed as relativistic generalisations of CMS systems and are closely connected with relativistic field theories such as the sine-Gordon model. The original relativistic CMS systems, often referred to as Ruijsenaars(--Schneider) systems, are again naturally associated with $A_{N-1}$ type root systems.

While the classical integrable CMS systems are given by $N$ Poisson commuting integrals that are polynomials in particle momenta, their generalisations to the relativistic level depend exponentially on generalised momenta. In the quantum case, we are thus dealing with (analytic) difference operators rather than partial differential operators, as is the case for nonrelativistic CMS systems.

Difference operators of a form similar to those considered by Ruijsenaars and associated with an arbitrary root system were introduced and studied by Macdonald (2000) in his fundamental work on multivariable $q$-polynomials.\footnote{The results in question were presented in a widely circulated 1987 preprint, which, however, was not published until 2000.}
In the $A_{N-1}$-case, Macdonald's difference operators are related to Ruijsenaars' original operators, which feature elliptic potentials, by degeneration to the trigonometric level and a similarity transformation. Further early related results, involving remarkable multivariable $q$-polynomials and relativistic CMS type systems associated with the non-reduced $BC_N$ root systems, are due to Koornwinder (1992) and van Diejen (1994).

CMS systems, both at the classical and quantum as well as the nonrelativistic and relativistic level, allow for a number of further generalisations preserving integrability. Examples include systems featuring internal degrees of freedom, supersymmetry as well as systems associated with very particular arrangements of hyperplanes with prescribed multiplicities that lack the reflection symmetry of a root system. In addition, numerous other types of (finite-dimensional) integrable systems can be obtained as limits of CMS systems.

Within the confines of this article, we cannot hope to review all of these different types of systems of CMS type, not to mention the countless applications to and connections with other areas within mathematics and physics. Our aim is rather more modest: To acquaint the reader with CMS systems, both classical and quantum as well as nonrelativistic and relativistic, and indicate, mainly through examples, some of their most important and remarkable properties, such as integrability, explicit solutions and dualities relating the different systems. For simplicity, we shall focus on systems of $A_{N-1}$-type and leave out internal degrees of freedom.

Before turning to the details, we briefly indicate further important connections and developments involving CMS systems that are not discussed in the main part of the text.

On all levels of CMS systems, suitable limit transitions lead to Toda type systems, which feature exponential nearest-neighbour interactions, with the original nonrelativistic systems discovered by Toda (1967) when searching for lattice models admitting explicit solutions. Quantum systems of $N$ nonrelativistic bosons on the line or the circle, interacting pairwise via a $\delta$-function potential, can be obtained as limiting cases of quantum nonrelativistic CMS systems. One can also use analytic continuation to produce systems consisting of more than one species of particles.

In fact, starting from CMS systems associated with root systems, or even more general configurations of vectors compatible with integrability, a large number of integrable many-body systems have been produced using appropriate limits as well as analytic continuation. We note that such transitions between different integrable quantum many-body systems are often established on the level of commuting Hamiltonians, where they tend to be relatively straightforward to handle. In contrast, on the level of solutions, both analytic continuations and limit transitions can be very challenging to control and many cases remain to be handled.

Another type of limit transition, that has been used to great effect to study integrable spin chains of Haldane--Shastry and Inozemtsev type, is the so-called freezing trick. Starting from a CMS system with spin degrees of freedom, the particles
are effectively ``frozen''  at equilibrium positions.

From an early stage, fundamental connections between CMS systems and numerous infinite-dimensional integrable field theories have played an important role in the development of the subject. Specific examples include pole dynamics of solutions to the Korteweg--de Vries and Kadomtsev--Petviashvili equations governed by a classical CMS system and soliton scattering in the sine-Gordon model reproduced by a relativistic CMS system, both at the classical and the quantum level.

Further examples of developments in theoretical physics for which CMS systems have turned out to be relevant can be found in subfields such as (supersymmetric) quantum field and string theories, quantum chaos, exactly solved models in statistical mechanics, the quantum Hall effect and more.

The interplay between CMS systems and numerous subfields of mathematics should also be highlighted. On the one hand, many different mathematical objects and tools have been used to construct and study CMS systems. On the other, there are various examples of mathematical developments in which CMS systems have played an important role. Specific examples of relevant subfields include symplectic- and algebraic geometry, the (representation) theory of Lie algebras and groups, quantum groups and double affine Hecke algebras, harmonic analysis on symmetric spaces, spectral theory, special functions, symmetric functions, random matrix theory, dynamical systems, etc.

Finally, we note that due, in particular, to the limited scope of this paper, our list of references only represents a small fraction of the vast literature on and related to CMS systems. With this in mind, we have compiled a list of suggested further reading, which we hope will help the interested reader to gain a broader perspective on CMS systems and some of the many related developments indicated above.

\section{Classical CMS systems}
A system of $N$ point particles of equal mass $m>0$, moving on the line and interacting pairwise can be described by a Hamiltonian of the form
\begin{equation}
\label{H}
H = \frac{1}{2m}\sum_{i=1}^N p_i^2+
\frac{g^2}{m}\sum_{1\leq i<j\leq N}V(x_i-x_j),
\end{equation}
with $x_i$ and $p_i$ the position and momentum of particle number $i$, respectively, and where $g$ is a coupling constant. The CMS systems correspond to very specific choices of pair potential $V$. There are four distinct cases: rational, hyperbolic, trigonometric and elliptic, commonly referred to as types $\text{I}-\text{IV}$.

In the rational or type $\text{I}$ case, we have
\begin{equation}
\label{VRat}
V(x) = \frac{1}{x^2}\ \ (\mathrm{I}).
\end{equation}
Assuming that $g^2>0$, the potential $(g^2/m)V(x_i-x_j)$ is repulsive and singular along $x_i=x_j$. Since the Hamiltonian is a constant of motion, it follows that the particle distance $|x_i-x_j|$ is bounded from below. Hence, the ordering of the particles remains unchanged as the system evolves in time, so that we may assume that $x_i>x_j$ for $i<j$, which means that the configuration space becomes the cone given by the inequalities
\begin{equation}
\label{RatConfig}
x_i-x_{i+1} > 0,\ \ \ i = 1,\ldots,N-1.
\end{equation}

The type $\text{II}$ case is characterised by the hyperbolic pair potential
\begin{equation}
\label{VHyp}
V(x) = \frac{a^2}{4\sinh^2(ax/2)}\ \ a>0\ \ (\mathrm{II}).
\end{equation}
It is clear that the rational case is recovered in the limit $a\to 0$ and that we again can take as configuration space the cone given by \eqref{RatConfig}.

Considering the $N$ particles as located on a circle of radius $1/a$ at angles $ax_i$, the chord-distance between particles number $i$ and $j$ in the plane of the circle is given by $2a^{-1}|\sin(a(x_i-x_j)/2)|$. Assuming pairwise interaction of the inverse-square of distance type, we are thus led to the trigonometric pair potential
\begin{equation}
\label{VTrig}
V(x) = \frac{a^2}{4\sin^2(ax/2)}\ \ (\mathrm{III}).
\end{equation}
Thanks to the identity
$$
\sum_{n\in\mathbb{Z}}\frac{1}{(x-2\pi n)^2} = \frac{1}{4\sin^2(x/2)},
$$
we can also think of the particles as located on a line with each particle periodically repeated and interacting pairwise with all copies of the remaining particles
through a potential proportional to the inverse-square of the distance between the particles.

The latter interpretation corresponds to the configuration space being the convex polytope defined by
$$
x_i-x_{i+1} > 0,\ \ i = 1,\ldots,N-1,\ \ \ x_1-x_N < 2\pi/a.
$$
If, on the other hand, we consider the particles as moving on a circle, then we should divide out by a suitable discrete group action. There are two possibilities: either we think of the particles as indistinguishable or as distinguishable. The relevant group action is generated by the map taking $(x_1,\ldots,x_N)$ to $(x_N+2\pi/a,x_1,\ldots,x_{N-1})$ or $(x_1+2\pi/a,\ldots,x_N+2\pi/a)$, respectively.

At the top level, we have the elliptic case
\begin{equation}
\label{VEll}
V(x) = \wp(x;\omega_1,\omega_2),\ \ \omega_1 > 0,\, -\ii \omega_2 > 0\ \ (\mathrm{IV}),
\end{equation}
where the Weierstrass $\wp$-function $\wp(x;\omega_1,\omega_2)$ is a meromorphic doubly-periodic function with periods $2\omega_1$ and $2\omega_2$ whose poles are of second order and located at $2(n\omega_1+m\omega_2)$, $n,m\in\mathbb{Z}$. The previous potential functions $\text{I}-\text{III}$ can all be obtained as limiting cases of \eqref{VEll}. Specifically, setting $\omega_1=\pi/a$ and taking $-\ii\omega_2\to\infty$, we obtain \eqref{VTrig}; whereas choosing $\omega_2=\ii\pi/a$ and sending $\omega_1\to\infty$ leads to \eqref{VHyp}. (In both cases up to an additive constant). Letting both periods go to infinity, one obtains \eqref{VRat}.

The above remarks on the configuration space of the type $\text{III}$ system applies also to the type $\text{IV}$ system, with the real period $2\omega_1$ of $\wp(x;\omega_1,\omega_2)$ replacing $2\pi/a$.

In the rational case, Moser (1975) proved that the CMS system is (Liouville) integrable, in the sense that it possesses $N$ independent integrals of motion $H_r=H_r(x,p)$, $r=1,\ldots,N$, globally defined and in involution, that is they Poisson commute. Here, independence refers to linear independence of the differentials $dH_r$. More specifically, he used Lax's method and observed that if the particle positions and momenta evolve in time according to Hamilton's equations
$$
\frac{dx_i}{dt} = \frac{\partial H}{\partial p_i},\ \ \ \frac{dp_i}{dt} = -\frac{\partial H}{\partial x_i},
$$
with the rational CMS Hamiltonian $H$, given by \eqref{H}--\eqref{VRat}, then the Lax equation
\begin{equation}
\label{LEq}
\frac{dL}{dt} = ML-LM
\end{equation}
is satisfied by the pair of $N\times N$ matrix-valued functions $L=(L_{ij})$ and $M=(M_{ij})$ (the so-called Lax pair) with matrix elements
\begin{equation}
\label{L}
L_{ij} = p_i\delta_{ij}+\ii g\frac{1-\delta_{ij}}{x_i-x_j},
\end{equation}
\begin{equation}
\label{M}
M_{ij} = \frac{\ii g}{m}\left(-\delta_{ij}\sum_{k\neq i}\frac{1}{(x_i-x_k)^2}+\frac{1-\delta_{ij}}{(x_i-x_j)^2}\right),
\end{equation}
where the Kronecker delta $\delta_{ij}$ equals $0$ if $i\neq j$ and $1$ if $i=j$. (As a direct computation reveals, the Lax equation amounts to Hamilton's equations for the particle differences $x_i-x_j$ and particle momenta $p_i$.)

It is now easy to see that the power traces
$$
H_r := \frac{1}{r}\mathrm{tr}(L^r),\ \ \ r = 1,\ldots,N
$$
are conserved quantities. Indeed, from \eqref{LEq} and the cyclic property of the trace follows that
$$
\frac{d}{dt}H_r = \mathrm{tr}\big((ML-LM)L^{r-1}\big) = \mathrm{tr}\big(ML^r\big)-\mathrm{tr}\big(LML^{r-1}\big)
= 0.
$$
We note that the first two power traces yield the total momentum and the Hamiltonian:
$$
H_1 = P := \sum_{i=1}^N p_i,\ \ \ H_2 = mH.
$$
Since the characteristic polynomial of $L$ is given by the expansion
$$
\det(\lambda\cdot 1_N-L) = \sum_{r=0}^N \lambda^{N-r}(-1)^rS_r,
$$
where the elementary symmetric functions $S_r$ of the eigenvalues of $L$ can be expressed, using Newton's identites, as polynomials in the power-traces $H_r$, it is  in fact the whole spectrum of $L(t)$ that is independent of time $t$.

To establish involutivity of the power traces, one can use the fact that $|x_i(t)-x_j(t)|\to\infty$ as $t\to\pm\infty$ for any initial data $(x_0,p_0)$ and $i\neq j$, which is natural to  expect given the repulsive character of the type $\text{I}$ potential. Specifically, by the Jacobi identity, the Poisson brackets
$$
G_{rs} := \{H_r,H_s\} = \sum_{i=1}^N\left(\frac{\partial H_r}{\partial x_i}\frac{\partial H_s}{\partial p_i}-\frac{\partial H_r}{\partial p_i}\frac{\partial H_s}{\partial x_i}\right)
$$
are integrals of motion, so that
$$
G_{rs}(x_0,p_0) = G_{rs}(x(t),p(t)),\ \ \ \forall t\in\mathbb{R}.
$$
Since each term in $G_{rs}$ contains at least one factor $(x_i-x_j)^{-1}$, we find in the limit $t\to\infty$ that $G_{rs}(x_0,p_0)=0$.

In the type $\text{II}$ case, integrability can be established using similar arguments and, given that the systems of type $\text{III}$ are obtained from the type $\text{II}$ systems by the substitution $a\mapsto \ii a$, the type $\text{III}$ systems are also seen to be integrable.

For the type $\text{IV}$ case, the Lax matrix $L(\lambda)$ introduced by Krichever (1980), which depends on an additional parameter $\lambda$, as well as the corresponding spectral curve $\det(\mu\cdot 1_N-L(\lambda))=0$, plays an important role. In particular, it can be used to obtain integrals of motion.

\section{Quantum CMS systems}
\label{Sec:3}
The quantisation of a classical many-body system with Hamiltonian \eqref{H} can be accomplished by the usual canonical quantization substitutions
\begin{equation}
\label{canQuant}
p_i\mapsto -\ii \hbar\frac{\partial}{\partial x_i},\ \ \ i  = 1,\ldots,N
\end{equation}
with Planck's constant $\hbar>0$. In combination with the substitution $g^2\to g(g-\hbar)$, which is convenient in the case of CMS systems, this leads to the Schr\"odinger operator
\begin{equation}
\label{qH}
\hat{H} = -\frac{\hbar^2}{2m}\sum_{i=1}^N\frac{\partial^2}{\partial x_i^2} + \frac{g(g-\hbar)}{m}\sum_{1\leq i<j\leq N}V(x_i-x_j).
\end{equation}
As will become clear below, this change in the $g$-dependence of $\hat{H}$ ensures that corresponding eigenfunctions and quantum Lax pairs depend on $g$ in a particularly simple way.

In analogy with the classical notion of integrability, a Schr\"odinger operator of the form \eqref{qH} will be called {\em integrable} if there exists $N$ (algebraically) independent and pairwise commuting partial differential operators (PDOs) $\hat{H}_r$, $r=1,\ldots,N$ that commute with $\hat{H}$. The PDOs $\hat{H}_r$ are then referred to as {\em quantum integrals} of $\hat{H}$. Here, requiring independence means that the operators in question should not satisfy any nontrivial polynomial relations.

As it turns out, producing quantum integrals of CMS systems is not simply a matter of applying the canonical quantization substitutions \eqref{canQuant} to classical integrals. A concrete example of this fact is provided by the construction due to Ujino, Hikami and Wadati (1992) of quantum analogs of the power traces $H_r=r^{-1}\text{tr}(L^r)$ of the type $\text{I}$ classical Lax matrix $L$. Their starting point is the quantum Lax pair $\hat{L}$, $\hat{M}$ obtained from the classical Lax pair $L$, $M$ by the canonical quantization substitutions \eqref{canQuant} and multiplication of $M$ by $i\hbar$. By a direct computation, the quantum Lax equation
$$
\big[\hat{L},\hat{H}\cdot 1_N\big] = \big[\hat{M},\hat{L}\big]
$$
is readily verified. Furthermore, using the fact that the equality remains valid after a substitution $\hat{L}\to\hat{L}^r$, $r\in\mathbb{Z}_{>0}$, as well as $\sum_{i=1}^N\hat{M}_{ij}=\sum_{j=1}^N\hat{M}_{ij}=0$, it is easily inferred that the {\em total} power-traces
\begin{equation}
\label{hatHr}
\hat{H}_r := r^{-1}\sum_{i,j=1}^N(\hat{L}^r)_{ij},\ \ \ r = 1,\ldots,N
\end{equation}
are quantum integrals of $\hat{H}$ with $V$ given by \eqref{VRat}. In the type $\text{II}-\text{III}$ cases analogous claims hold true.

For earlier approaches towards constructing quantum integrals of CMS systems and, in particular, the subtleties involved in obtaining quantum versions of the symmetric functions $S_r$ of $L$, see e.g.~the reviews by Olshanetsky and Perelomov (1983) and by Ruijsenaars (1994).

Integrability of quantum CMS systems can also be established using Dunkl differential-difference operators. First introduced by Dunkl (1989) in the context of special functions associated with reflection groups, they have become an indispensable tool for the construction and study of quantum integrable systems of CMS type.

Let $S_N$ denote the symmetric group in $N$ elements, which, in particular, is generated by the transpositions $\sigma_{ij}$, $1\leq i<j\leq N$, which act on a function $f(x)$ of $x=(x_1,\ldots,x_N)$ by exchanging variables $x_i$ and $x_j$. In the rational case, Dunkl operators are given by
$$
D_i(k) = \frac{\partial}{\partial x_i}+k\sum_{j\neq i}\frac{1}{x_i-x_j}(1-\sigma_{ij}),\ \ \ i = 1,\ldots,N,
$$
where $k$ is a (complex) parameter. Their most remarkable property is pairwise commutativity:
$$
[D_i,D_j] = 0,\ \ \ \forall i,j = 1,\ldots,N.
$$
Further important properties include $S_N$-equivariance, which amounts to
$$
\sigma_{ij}D_i\sigma_{ij} = D_j,\ \ \sigma_{jl}D_i\sigma_{jl} = D_i,\ \ \ i\neq j,l;
$$
and the fact that $D_i$ leaves the space of (complex) polynomials $\mathbb{C}[x_1,\ldots,x_N]$ in the variables $x_1,\ldots,x_N$ invariant.

Thanks to the $S_N$-equivariance of Dunkl operators, the k-Laplacian
$$
\Delta_k := \sum_{i=1}^ND_i(k)^2 = \sum_{i=1}^N\frac{\partial^2}{\partial x_i^2}+2k\sum_{1\leq i<j\leq N}\frac{1}{x_i-x_j}\left(\frac{\partial}{\partial x_i}-\frac{\partial}{\partial x_j}-\frac{1-\sigma_{ij}}{x_i-x_j}\right)
$$
commutes with each transposition $\sigma_{ij}$, i.e.~$\sigma_{ij}\Delta_k=\Delta_k\sigma_{ij}$. Hence, it can be restricted to act on the subspace $\Lambda_N\subset\mathbb{C}[x_1,\ldots,x_N]$ of symmetric polynomials $p$, which, by definition, satisfy $p(\sigma_{ij}x)=p(x)$. The resulting operator is clearly given by
$$
\Delta_k\big\arrowvert_{\Lambda_N} = \sum_{i=1}^N\frac{\partial^2}{\partial x_i^2}+2k\sum_{1\leq i<j\leq N}\frac{1}{x_i-x_j}\left(\frac{\partial}{\partial x_i}-\frac{\partial}{\partial x_j}\right).
$$
As can be verified by a direct computation, it is essentially gauge-equivalent to the type $\text{I}$ CMS Schr\"odinger operator:
$$
\hat{H}(g) = -\frac{\hbar^2}{2m}W(x;g)^{1/2}\Delta_{g/\hbar}\big\arrowvert_{\Lambda_N}W(x;g)^{-1/2}
$$
with the positive weight function
$$
W(x;g) = \left(\prod_{1\leq i<j\leq N}(x_i-x_j)^2\right)^{g/\hbar},
$$
and where $V$ in $\hat{H}$ is given by \eqref{VRat}.

More generally, any $p\in\Lambda_N$ gives rise to a PDO
\begin{equation}
\label{hHp}
\hat{H}_p(g) := W(x;g)^{1/2}p(-\ii\hbar D_1(g/\hbar),\ldots,-\ii\hbar D_N(g/\hbar))\big\arrowvert_{\Lambda_N}W(x;g)^{-1/2}
\end{equation}
and, from the pairwise commutativity of Dunkl operators, it follows immediately that $[\hat{H}_p,\hat{H}_q]=0$ for all $p,q\in\Lambda_N$. In particular, taking $p(x)=r^{-1}p_r(x)$, with the power sum symmetric functions
$$
p_r(x) := x_1^r+\cdots+x_N^r,\ \ \ r=1,\ldots,N,
$$
the quantum integrals $\hat{H}_r$ \eqref{hatHr} are recovered.

Applying the above change of gauge to the Dunkl operators themselves yields
$$
W(x;g)^{1/2}(-\ii\hbar D_i(g/\hbar))W(x;g)^{-1/2} = -\ii\hbar \frac{\partial}{\partial x_i}+\ii g\sum_{j\neq i}\frac{1}{x_i-x_j}\sigma_{ij} = \pi_i(g),
$$
with $\pi_i$ the `coupled' momentum operators introduced by Polychronakos (1992). On this level, there is a clear similarity between the Dunkl operators and the (quantum) Lax matrix. This connection has been explored in great detail and generality by Chalykh (2019).

In the type $\text{II}$-$\text{III}$ cases, one can use the global Dunkl operators introduced by Heckman (1991), trigonometric analogues of Polychronakos operators or the trigonometric Dunkl operators defined by Cherednik (1991). These operators have slightly different properties, with the first two being equivariant while lacking pairwise commutativity, whereas Cherednik's operators do commute but are not equivariant.

Elliptic Dunkl operators were introduced by Buchstaber, Felder and Veselov (1994). Since they, like Krichever's Lax matrix, depend on an additional parameter, it is a delicate matter to obtain quantum integrals of the CMS system with the type $\text{IV}$ pair potential \eqref{VEll}. This problem was resolved by Etingof, Felder, Ma and Veselov (2011).

\section{Explicit solutions}

We proceed to indicate the nature of solutions of CMS systems and provide a few examples of how they can be constructed. To simplify the exposition, we set the mass $m=1$ and, in the quantum case, Planck's constant $\hbar=1$. By suitably scaling model parameters, both $m$ and $\hbar$ are easily reintroduced.

\subsection{The classical level}
The Liouville--Arnold theorem guarantees that Hamilton's equations of motion for an integrable system can be solved by quadrature, i.e.~its solutions can be expressed in terms of integrals. However, the theorem is of a qualitative nature and offers little help with extracting concrete information on the solutions.

It is therefore a remarkable feature of CMS systems that explicit expressions for their solutions have been obtained. For example, the projection method of Olshanetsky and Perelomov (1976) identifies the systems as projections of higher-dimensional systems with equations of motion that are easily integrated. 

More precisely, it is readily verified that the type $\text{I}$ Lax pair $L,M$, as given by \eqref{L}--\eqref{M}, satisfies the equation
\begin{equation}
\label{LPMX}
L = P-[M,X],\ \ \ X = \text{diag}(x_1,\ldots,x_N),\ \ P = \dot{X}.
\end{equation}
Let $u(t)$ be the solution of $\dot{u}=-uM$ such that $u(0)=1_N$. Then, a direct computation using $du^{-1}/dt=-u^{-1}\dot{u}u^{-1}$, the Lax equation \eqref{LEq} and \eqref{LPMX} reveals that
$$
Q(t) := u(t)X(t)u(t)^{-1}
$$
satisfies $\dot{Q}=uLu^{-1}$ and $\ddot{Q}=0$. It follows that
\begin{equation}
\label{QExpl}
Q(t) = X(0)+tL(0)
\end{equation}
and the particle positions $x_i(t)$ can be obtained as the eigenvalues of $Q(t)$. In effect, the problem of solving the equations of motion for the CMS system of type $\text{I}$ is reduced to the {\em algebraic} problem of computing the eigenvalues of the matrix $Q(t)$, given explicitly by \eqref{L} and \eqref{QExpl}.

In other words, the solution of the CMS system of type $\text{I}$ can be obtained from geodesic flow (i.e.~free motion) in the space of $N\times N$ hermitian matrices. A conceptual explanation of this result can be found within the Hamiltonian reduction approach to CMS systems, as developed by Kazhdan, Kostant and Sternberg (1978).

In the type $\text{II}$ and $\text{III}$ cases, one can instead exploit geodesic flow in spaces of positive definite hermitian and unitary matrices, respectively; see e.g.~the book by Perelomov (1990) and references therein.

The type $\text{IV}$ system is more difficult to handle and information one can obtain is typically less explicit. An early important result on its solutions is due to Krichever (1980), who showed that the corresponding equations of motion can be integrated in terms of a Riemann theta function.

\subsection{The quantum level}
\label{Sec:42}
While integrability in itself yields little concrete information on solutions, a great deal is known about joint eigenfunctions of the quantum integrals of the CMS operator $\hat{H}$ \eqref{qH}. In contrast to the classical case, the specific value of the coupling constant $g$ can have a significant impact on the complexity of the solutions. 

To illustrate this fact, let us first consider the type $\text{I}$ case for
\begin{equation}
\label{BAVals}
g = -m,\ \ \ m\in\mathbb{Z}_{>0}.
\end{equation}
As first observed by Chalykh and Veselov (1990), there exists a function of the form
\begin{equation}
\label{psi}
\psi(\lambda,x) = P(\lambda,x)e^{\ii \lambda\cdot x},\ \ \ \lambda,x\in\mathbb{C}^N
\end{equation}
with the joint eigenfunction property
$$
\hat{H}_r(x)\psi(\lambda,x) = r^{-1}p_r(\lambda)\psi(\lambda,x), \ \ \ r = 1,\ldots,N,
$$
and where $P(\lambda,x)$ is a rational function in $x$ and a polynomial in $\lambda$ with highest degree term $A_m(\ii\lambda)$, where
$$
A_m(x) = \prod_{1\leq i<j\leq N}(x_i-x_j)^m.
$$
This so-called rational Baker-Akhiezer (BA) function is given explicitly by
\begin{equation}
\label{Berest}
\psi(\lambda,x) = \frac{1}{M!}\big(2^{-1}p_2(\lambda)-\hat{H}(x)\big)^M\big(A_m(x) e^{\ii\lambda\cdot x}\big),\ \ \ M = mN(N-1)/2,
\end{equation}
a formula which is due to Berest (1998).

From the particular form \eqref{VRat} of the pair-potential in the type $\text{I}$ case, it is readily inferred that a meromorphic solution of $\hat{H}\psi=2^{-1}p_2(\lambda)\psi$ has either a pole of order $m$ or a zero of order $m+1$ along each hyperplane $x_i=x_j$, $1\leq i<j\leq N$. For the BA function, the expression \eqref{Berest} clearly implies the former, whereas the transformation property $\psi(\sigma_{ij}\lambda,x)=(-1)^m\psi(\lambda,\sigma_{ij}x)$ entails that the latter is the case for its antisymmetrisation. In other words, we have
$$
\Psi(\lambda,x) := \sum_{\sigma\in S_N}(-1)^{|\sigma|}\psi(\sigma\lambda,x) = A_{m+1}(x)J_m(\lambda,x)
$$
for some analytic $S_N$-invariant function $J_m(\lambda,x)$ such that $J_m(\lambda,0)\neq 0$.

The function $J_m$ essentially amounts to the $A_{N-1}$-instance of the multivariable Bessel functions associated with root systems, as introduced by Opdam (1993). In contrast to the BA function, the $A_{N-1}$ Bessel function is defined and provides regular `physical' solutions to the rational CMS system whenever the coupling constant $g^2\geq 0$. Although it is not of the elementary form \eqref{psi} for generic parameter values, it has a series representation that is explicit in terms of Jack polynomials (see e.g.~R\"osler (2003)) as well as a representation by multidimensional integrals whose integrands are elementary functions.

To indicate the physical significance of the joint eigenfunction $\psi(\lambda,x)$, we take $\lambda=p$ with $p=(p_1,\ldots,p_N)\in\mathbb{R}^N$ representing particle momenta. From Berest's formula, it is straightforward to infer the dominant asymptotic behaviour
$$
\Psi(p,x)\sim \sum_{\sigma\in S_N}(-1)^{|\sigma|}A_m(\ii \sigma p)e^{\ii\sigma p\cdot x}
$$
for $x_i-x_{i+1}\to\infty$, $i=1,\ldots,N-1$. This entails that the particles in the type $\text{I}$ CMS system exhibit soliton scattering (conservation of momenta and factorisation of the $S$-matrix). Indeed, let us take $p_1<p_2<\cdots<p_N$, so that in the region $x_1>x_2>\cdots>x_N$ there is a unique incoming wave $A_m(\ii p)e^{\ii p\cdot x}$ as well as a unique outgoing wave $(-1)^{|\sigma_0|}A_m(\ii \sigma_0 p)e^{\ii \sigma_0p\cdot x}$, associated with the order reversing permutation
$$
\sigma_0(i) = N+1-i,\ \ \ i = 1,\ldots,N.
$$
We note that, since only terms of the form $e^{\ii \sigma p\cdot x}$ occur, the momenta $p_1,\ldots,p_N$ are conserved throughout the scattering process; and the corresponding $S$-matrix element is a momentum-independent phase, given by
$$
S = \frac{(-1)^{|\sigma_0|}A_m(\ii \sigma_0 p)}{A_m(\ii p)} = (-1)^{(1-m)N(N-1)/2}.
$$

In the type $\text{II}$ case, a so-called trigonometric BA function exists for the parameter values \eqref{BAVals}, whereas the arbitrary-$g$ joint eigenfunctions are given by a multivariable hypergeometric function introduced and studied by Heckman and Opdam (1987). To be precise, they associate a hypergeometric function to an arbitrary root system and the present case corresponds to the root system $A_{N-1}$. From their construction, one can read off the factorised asymptotics of the eigenfunctions and the corresponding $S$-matrix element takes the form
$$
S(p) = \prod_{1\leq i<j\leq N}\big(-u(p_j-p_i)\big),\ \ \ u(v) = \frac{\Gamma(\ii v+1)\Gamma(-\ii v+g)}{\Gamma(-\ii v+1)\Gamma(\ii v+g)}.
$$
We note that $S(p)$, again a phase, factorises into $N(N-1)/2$ two-particle elements that, in contrast to the type $\text{I}$ case, do depend on the momenta $p_1,\ldots,p_N$.

Unlike in the type $\text{I}$ and type $\text{II}$ cases, in which the physically relevant eigenfunctions are transcendental for generic $g$-values, the type $\text{III}$ eigenvalue problem is essentially polynomial in nature. Specifically, for each partition $\lambda=(\lambda_1,\ldots,\lambda_N)$, i.e.~$\lambda\in\mathbb{Z}_{\geq 0}^N$ and $\lambda_1\geq\cdots\geq\lambda_N$, there is a joint eigenfunction of the form
\begin{equation}
\label{Psilam}
\Psi_\lambda(x) = W(x)^{1/2}P_\lambda(x)
\end{equation}
with positive weight function
$$
W(x) = \left(\prod_{1\leq i<j\leq N}4\sin^2(a(x_i-x_j)/2)\right)^g,
$$
and where $P_\lambda$ denotes the symmetric Jack polynomials, which first arose as a one-parameter generalisation of the Schur polynomials in highly influential work by Jack (1970). We note that the corresponding joint eigenvalues are polynomials in the quasi-momenta $a\lambda_i+ag(N-2i+1)/2$, $i=1,\ldots,N$; and that $W(x)^{1/2}$ is the groundstate wave function of the type $\text{III}$ CMS system.

The type $\text{IV}$ case is, in many ways, the most challenging to study, but a number of explicit results on joint eigenfunctions have nevertheless been obtained. In particular, Felder and Varchenko (1997) obtained eigenfunctions of so-called Bethe ansatz type when the coupling parameter $g=-m$, $m\in\mathbb{Z}_{>0}$. In the $N=2$ case, the Schr\"odinger equation amounts to the Lam\'e equation and solutions with a Bethe ansatz structure date back to work by Hermite. For general $g$-values, perturbative constructions of eigenfunctions were established by Komori and Takemura (2002) as well as Langmann (2014).

It is also worth noting that, at the time of writing, elliptic CMS systems, their eigenfunctions and related problems, in particular in algebraic geometry and the theory of (supersymmetric) gauge theories, play a central role in several research activities. To get a feel for these activities or gain a broader perspective on earlier work on the type $\text{IV}$ systems, we can recommend the review by Ruijsenaars (2004), the proceedings from the 2019 Nordita workshop Elliptic integrable systems, special functions and quantum field theory (Halln\"as, Noumi, Spiridonov and Warnaar, 2020) as well as the recorded lectures from Koroteev's 2021 and 2022 workshops Elliptic integrable systems, available at https://math.berkeley.edu/$\sim$pkoroteev/ workshop2.html (accessed on 27 June 2023).

\section{Relativistic Calogero--Moser--Sutherland systems}

For each of the types $\text{I}-\text{IV}$ of pair potential $V$, Ruijsenaars and Schneider (1986) in the classical- and Ruijsenaars (1987) in the quantum case introduced and studied an integrable one-parameter deformation of the corresponding CMS system. The `extra' parameter has a natural interpretation in terms of the speed of light $c$ and, after subtracting the rest energy $Nmc^2$, the defining Hamiltonian reduces to its nonrelativistic counterpart in the limit $c\to\infty$.

\subsection{The classical level}
The nonrelativistic Hamiltonian \eqref{H} and total momentum $P=\sum_{i=1}^Np_i$ are replaced by functions of the form
\begin{equation}
\label{Hrel}
H = \frac{1}{m\beta^2}\sum_{i=1}^N \cosh(\beta p_i)\prod_{j\neq i}f(x_i-x_j),
\end{equation}
\begin{equation}
P = \frac{1}{\beta}\sum_{i=1}^N \sinh(\beta p_i)\prod_{j\neq i}f(x_i-x_j),
\end{equation}
where it is natural to take the deformation parameter $\beta=1/mc>0$ with $c$ the speed of light.

In the special case $f=1$, they describe $N$ free relativistic particles with mass $m$ in terms of their rapidities $\beta p_i$ and, together with the (Lorentz) boost generator
$$
B = -m\sum_{i=1}^N x_i,
$$
they provide a representation of the Poincar\'e Lie algebra
$$
\{H,P\} = 0,\ \ \{H,B\} = P,\ \ \{P,B\} = m^2\beta^2 H.
$$
The last two equalities clearly hold true for any choice of function $f$. In contrast, as soon as $N>2$, the Poisson commutativity of $H$ and $P$ amounts to a nontrivial functional equation for $f$, which is satisfied whenever
$$
f^2(x) = a+b\wp(x),\ \ \ a,b\in\mathbb{C}.
$$
Taking $a,b>0$ and the periods $\omega_1,\omega_2$ of $\wp$ as in \eqref{VEll}, the right-hand side becomes a positive function for real $x$ and we can take the positive square-root to obtain the type $\text{IV}$ or elliptic relativistic CMS system. As in the nonrelativistic case, sending either one or both of $\omega_1,-\ii\omega_2$ to $\infty$ leads to the type $\text{I}-\text{III}$ cases. Specifically, it is convenient to work with the functions
$$
f(x) =
\left\{
\begin{array}{ll}
\big(1+(g\beta)^2/x^2)^{1/2} & (\text{I}\big)\\
\big(1+\sin^2(ag\beta/2)/\sinh^2(ax/2)\big)^{1/2} & (\text{II})\\
\big(1+\sinh^2(ag\beta/2)/\sin^2(ax/2)\big)^{1/2} & (\text{III})
\end{array}
\right.
$$
where the radicand is manifestly positive.

A remarkable property of the relativistic CMS systems is the existence of integrals of motion given by the simple and explicit formula
\begin{equation}
\label{Spmk}
S_{\pm r} = \sum_{\substack{I\subset\{1,\ldots,N\}\\ |I|=r}} \exp\left(\pm\beta\sum_{i\in I} p_i\right)\prod_{i\in I,j\notin I}f(x_i-x_j)
\end{equation}
with $r=1,\ldots,N$, and where
$$
S_{-r} = S_{N-r}S_N^{-1},\ \ \ S_N = \exp(\beta(p_1+\cdots+p_N)).
$$
Note that $H=(S_1+S_{-1})/(2m\beta^2)$ and $P=(S_1-S_{-1})/(2\beta)$.

The integrals $S_1,\ldots,S_N$ amount to symmetric functions of a Lax matrix $\mathscr{L}=(\mathscr{L}_{ij})$. Specifically, starting with the type $\text{II}$ case, one can use suitable substitutions in Cauchy's identity
$$
\det\left(\frac{1}{z_i-w_j}\right)_{i,j=1}^N = \prod_{i=1}^N\frac{1}{z_i-w_i}\prod_{1\leq i<j\leq N}\frac{(z_i-z_j)(w_i-w_j)}{(z_i-w_j)(w_i-z_j)}
$$
to show that the principal minor $[\mathscr{L}]_{I,I}$, given by the rows and columns with index from $I\subset\{1,\ldots,N\}$, of the matrix
$$
\mathscr{L}_{ij} = d_iC_{ij}d_j,
$$
where
$$
d_i = \exp((ax_i+\beta p_i)/2)\prod_{j\neq i}f(x_i-x_j)^{1/2}
$$
and
$$
C_{ij} = \exp(-a(x_i+x_j)/2)\frac{\sinh(\ii\beta ag/2)}{\sinh(a(x_i-x_j+i\beta g)/2)},
$$
precisely match the corresponding term in \eqref{Spmk}. The substitution $a\to ia$ and limit $a\to 0$ yields matrix elements for the type $\text{III}$ and type $\text{I}$ case, respectively. In the type $\text{IV}$ case, an elliptic analogue of Cauchy's identity can be used, leading to a Lax matrix with spectral parameter.

The Lax matrices $L$ for the nonrelativistic CMS systems can be recovered as limiting cases:
$$
\mathscr{L} = 1_N+\beta L+O(\beta^2),\ \ \ \beta\to 0.
$$
Expanding the determinant of $\lambda\cdot 1_N-\beta^{-1}(\mathscr{L}-1_N)$ and taking $\beta\to 0$ yields a limit formula for the symmetric functions of $L$ in terms of the symmetric functions of $\mathscr{L}$. In this way, integrability of relativistic CMS systems entail integrability of corresponding nonrelativistic CMS systems.

\subsection{The quantum level}
\label{Sec:52}
In contrast to the nonrelativistic case, quantization of relativistic CMS systems involves nontrivial ordering problems even at the level of the defining Hamiltonian. Indeed, applying the canonical quantization substitutions \eqref{canQuant} to the classical Hamiltonian \eqref{Hrel} or, more generally, the integrals of motion \eqref{Spmk}, with the implied ordering of factors, yields operators that are neither (formally) self-adjoint nor commute.

However, there exists a factorisation $f(x)=f_+(x)f_-(x)$ such that the analytic difference operators (A$\Delta$Os)
$$
\hat{S}_{\pm r} := \sum_{\substack{I\subset\{1,\ldots,N\}\\|I|=r}}\prod_{i\in I,j\notin I}f_\mp(x_i-x_j) \exp\left(\mp \ii\hbar\beta\sum_{i\in I} \frac{\partial}{\partial x_i}\right)\prod_{i\in I,j\notin I}f_\pm(x_i-x_j)
$$
with $r=1,\ldots,N$ are self-adjoint and pairwise commute. We note that the operator exponentials $\exp(\mp i\hbar\beta\partial/\partial x_i)$ act on a function $F(x)$ that is analytic in $x_1,\ldots,x_N$ in a strip around $\mathbb{R}$ of width at least $2\hbar\beta$ by a complex shift of the argument:
$$
\exp(\mp \ii\hbar\beta\partial/\partial x_i)F(x)=F(x\mp \ii\hbar\beta e_i),
$$
where $e_i$ denotes the standard basis vectors in $\mathbb{R}^N$ given by $(e_i)_j=\delta_{ij}$. This motivates the A$\Delta$O-terminology.

For the elliptic type $\text{IV}$ case, Ruijsenaars (1987) identified such a factorisation in terms of Weierstrass $\sigma$-function and proved that integrability amounts to a sequence of functional identities for $\sigma$. In the limiting cases $\text{I}-\text{III}$, the factorisation is given by
$$
f_\pm(x) =
\left\{
\begin{array}{ll}
\big(1\pm \ii g\beta/x)^{1/2} & (\text{I}\big)\\
\big(\sinh(a(x\pm \ii g\beta)/2)/\sinh(ax/2)\big)^{1/2} & (\text{II})\\
\big(\sin(a(x\pm \ii g\beta)/2)/\sin(ax/2)\big)^{1/2} & (\text{III})
\end{array}
\right.
$$
(where the branch is fixed by $f_\pm(x)\to 1$ as $g\to 0$).

It is worth pointing out that, for a suitable weight function $W(x)$, the similarity transformed A$\Delta$Os $W(x)^{-1/2}\hat{S}_rW(x)^{1/2}$ feature meromorphic coefficients, which, from the point of view of eigenfunctions, often makes them easier to work with. In particular, working in such a meromorphic gauge, Chalykh (2000) showed that type $\text{I}$ joint eigenfunctions are given by the trigonometric Baker-Akhiezer function ($g=-m$ with $m\in\mathbb{Z}_{>0}$) and the $A_{N-1}$-version of the Heckman--Opdam hypergeometric function (general $g$), with the relevant rational difference operators acting in the spectral variables.

In the type $\text{II}$ case, joint eigenfunctions featuring factorised asymptotics were constructed and studied by Halln\"as and Ruijsenaars (2014), in particular confirming Ruijsenaars' conjecture that particles in the relativistic hyperbolic CMS system exhibit soliton scattering and that, for suitable values of the coupling parameter $g$, the soliton scattering in the sine-Gordon model is recovered.

Similarly to the nonrelativistic system, the type $\text{III}$ case features eigenfunctions of the form \eqref{Psilam} with a weight function that naturally factorises in terms of a trigonometric analogue of the Euler Gamma function, and where the relevant polynomials are Macdonald polynomials. To be precise, the polynomials in question are the symmetric $GL$-type polynomials introduced by Macdonald (1988), which can be viewed as natural $q$-analogs of the symmetric Jack polynomials.

Also the relativistic type $\text{IV}$ systems are, at the time of writing, receiving considerable attention in the literature and we can recommend the same references as in the nonrelativistic case.

\section{Action-angle and bispectral dualities}

The Liouville--Arnold theorem on integrable systems ensures the existence of a so-called action-angle map, a canonical transformation $\Phi$ that diagonalises all integrals of motion $I$ in the sense that $I\circ\Phi^{-1}$ is a function of the new generalized momenta (or action variables) only.
In the case of CMS systems, nonrelativistic as well as relativistic, such action-angle maps can be constructed explicitly and reveal remarkable action-angle dualities among different systems.

The quantum analogue of this picture is given by unitary joint eigenfunction transforms, with the spectral theorem playing the role of the Liouville--Arnold theorem. Specifically, for CMS systems, joint eigenfunctions considered in Sections \ref{Sec:42} and \ref{Sec:52} provide kernels for such transforms and the classical action-angle dualities correspond to bispectral properties of the joint eigenfunctions.

\subsection{The classical level}
In the nonrelativistic type $\text{I}$ case, an action-angle map, with generalized momenta equal to asymptotic momenta, was explicitly constructed by Airault, McKean and Moser (1977). For the type $\text{I}-\text{III}$ cases, both nonrelativistic and relativistic, explicit action-angle maps were obtained by Ruijsenaars (1988, 1995).

In the rational nonrelativistic case, his starting point is the readily verified commutation relation
$$
\frac{1}{\ii g}\lbrack A,L\rbrack = e\otimes e-1_N,
$$
with $A=\text{diag}(x_1,\ldots,x_N)$, $L$ the Lax matrix and $e=(1,\ldots,1)^t$. Since $L$ is hermitian, there exists a unitary matrix $U$ such that
$$
\tilde{L} := ULU^* = \text{diag}(\tilde{p}_1,\ldots,\tilde{p}_N).
$$
Applying $U$ to the above commutation relation and letting $\tilde{A}=UAU^*$, $\tilde{e}=Ue$ and $\check{e}^t=e^tU^*$, we obtain
$$
\frac{1}{\ii g}\tilde{A}_{ij}(\tilde{p}_j-\tilde{p}_i) = \tilde{e}_i\check{e}_j-\delta_{ij}.
$$
It follows that $\tilde{e}_i\check{e}_i=1$ (from $i=j$) and $\tilde{p}_j\neq\tilde{p}_i$ (from $i\neq j$), so that $U$ can be fixed uniquely by requiring that
$$
\tilde{p}_1 > \cdots > \tilde{p}_N,\ \ \ \tilde{e}_i = \check{e}_i = 1.
$$
Setting $\tilde{x}_i=\tilde{A}_{ii}$, we have thus constructed a map $\Phi:(x,p)\mapsto (\tilde{x},\tilde{p})$, with $\tilde{p}$ an element of the type $\text{I}$ configuration space, and the power traces $\tilde{H}_r:=H_r\circ\Phi^{-1}=r^{-1}\sum_{i=1}^N\tilde{p}_i^r$. Furthermore, from the above discussion it is clear that
$$
\tilde{A} = L(-g;\tilde{p},\tilde{x}),
$$
so that, in particular, the pull-backs $(J_r\circ\Phi^{-1})(\tilde{x},\tilde{p})$ of the power traces $J_r:=r^{-1}\mathrm{tr}(A^r)$ by $\Phi^{-1}$ yield an integral system that is again a CMS system of type $\text{I}$. In this sense, the type $\text{I}$ nonrelativistic CMS system is self-dual.

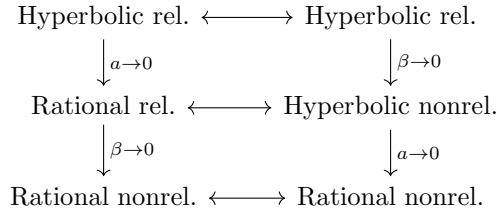
\begin{figure}[h]
$$
\begin{tikzcd}
\text{Hyperbolic rel.} \arrow[r,leftrightarrow] \arrow[d,"a\to 0"] & \text{Hyperbolic rel.} \arrow[d,"\beta\to 0"]\\
\text{Rational rel.} \arrow[r,leftrightarrow] \arrow[d,"\beta\to 0"] & \text{Hyperbolic nonrel.} \arrow[d,"a\to 0"]\\
\text{Rational nonrel.} \arrow[r,leftrightarrow] & \text{Rational nonrel.}
\end{tikzcd}
$$
\caption{Vertical arrows indicate limit transitions and horizontal arrows action-angle dualities.}
\label{Fig:dualtrans}
\end{figure}

Action-angle dualities along with limit transitions between systems of type $\text{I}$-$\text{II}$ are summarised in Figure \ref{Fig:dualtrans}. For details on the type $\text{III}$ case, we refer the reader to the original source listed above. From the point of view of hamiltonian reduction, duality relations were considered by Fock, Gorsky, Nekrasov and Rubtsov (2000) and studied in great detail by Feh\'er and coworkers, see e.g. Feh\'er and Marshall (2019) and references therein.

Systems dual to the (non)relativistic elliptic type $\text{IV}$ systems should exhibit an elliptic dependence on momenta and (rational) hyperbolic/trigonometric dependence on particle coordinates. Moreover, it is natural to expect that all of these systems arise as suitable limits of a self-dual integrable system featuring elliptic dependence on both particle momenta and coordinates. In the $N=2$ case, such a `double-elliptic' (DELL) system was constructed explicitly by Braden, Marshakov, Mironov and Morozov (2000). To do the same in the general-$N$ case remains an intriguing open problem.

\subsection{The quantum level}
From the explicit formulae recalled in Sections \ref{Sec:3} and \ref{Sec:52}, it is readily seen that quantum integrals of CMS systems are formally self-adjoint on $L^2(G,dx)$, where $G$ denotes the relevant configuration space. With the spectral theorem in mind, it is thus natural to expect that the quantum integrals can be simultaneously diagonalised by a unitary joint eigenfunction transform
$$
\mathcal{F}: L^2(G,dx)\to L^2(\hat{G},d\mu(p))
$$
of the form
$$
\mathcal{F}(\varphi)(p) = \int_G\Psi(-p,x)\varphi(x)dx.
$$
Here, diagonalisation of a quantum integral $\hat{H}_r$ refers to $\mathcal{F}(\hat{H}_r\varphi)(p)=m_r(p)\mathcal{F}(\varphi)(p)$ for some real-valued function $m_r$ (and appropriate $\varphi$); one can take $\hat{G}=G$ in the cases $\text{I}$-$\text{II}$, while in the type $\text{III}$-$\text{IV}$ cases $\hat{G}$ denotes a lattice encoding the joint spectrum; $d\mu$ is a suitable measure on $\hat{G}$; and the kernel $\Psi(p,x)$ is a joint eigenfunction of the relevant quantum integrals. 

In the nonrelativistic type $\text{I}$ case, the joint eigenfunction transform amounts to a symmetric analogue of the so-called Dunkl transform, which was studied in detail by Dunkl (1992) and de Jeu (1993) in the more general setting of an arbitrary root system. It can also be viewed as a multivariable Hankel transform, to which it essentially reduces in the $N=2$ case. When suitably normalised, the relevant joint eigenfunction, given by Opdam's multivariable Bessel function of $A_{N-1}$-type, is known to have the self-duality property $\Psi(p,x)=\Psi(x,p)$. As a consequence, it solves the bispectral problem
$$
\hat{H}_r(x)\Psi = m_r(p)\Psi,\ \ \ \hat{H}_r(p)\Psi = m_r(x)\Psi,
$$
with $\hat{H}_r$ given by \eqref{hatHr} and $m_r(z)=r^{-1}p_r(z)$. Furthermore,
in analogy with the classical case, type $\text{I}$ quantum integrals are diagonalised by both $\mathcal{F}$
as well as its inverse $\mathcal{F}^{-1}$.

The nonrelativistic type $\text{II}$ and relativistic type $\text{I}$ cases are covered by the $A_{N-1}$-instance of the hypergeometric Fourier transform, as developed by Opdam (1995). The eigenfunction transform for the relativistic type $\text{II}$ case is studied in work by Belousov, Derkachov, Kharchev and Khoroshkin (2023). In the type $\text{III}$ cases, the expected results on the eigenfunction transform readily follow from well-known orthogonality results for Jack- and Macdonald polynomials; see e.g.~the book by Macdonald (1995).

As in the classical case, an integrable system dual to the relativistic elliptic (type $\text{IV}$) CMS system and, more generally, the expected self-dual DELL system remains elusive. However, important and striking results in this direction have been obtained: the non-stationary Ruijsenaars functions defined by Shiraishi (2019), as well as their elliptic generalisations, are expected to provide eigenfunctions of such systems in the stationary limit; and the double-elliptic, but not self-dual, Hamiltonians introduced by Koroteev and Shakirov (2020) provide important insights into the nature of the desired DELL system. An overview of the state of the art in the subject can be found in the review by Mironov and Morozov (2023).

\section{Conclusions}

In this article, we have reviewed classical and quantum many-body systems of CMS type, both nonrelativistic and relativistic. For simplicity, we restricted attention to the original systems, which are naturally associated with root systems of type A, and highlighted some of their most fundamental properties, such as integrability, existence of exact solutions as well as important duality relations.

We hope that this article has provided the reader with a first glimpse of the remarkable nature of CMS systems and inspiration to delve deeper, e.g.~by consulting one or more of the papers or books suggested as further reading.

Finally, we would like to highlight the fact that research on and around CMS systems is very much alive and well. As previously indicated, an important open problem, attracting considerable interest from both mathematicians and physicists, is the search for the self-dual DELL system.  Another example of an intriguing open problem concerns Schr\"odinger operators of the form
$$
\hat{H}_{\mathcal{A}} = -\frac{1}{2}\sum_{i=1}^N \frac{\partial^2}{\partial x_i^2} + \sum_{\alpha\in\mathcal{A}}\frac{g_\alpha(g_\alpha-1)(\alpha,\alpha)}{2}V((\alpha,x)),
$$
given by a collection of nonparallel vectors $\mathcal{A}\subset\mathbb{R}^N$ and a `multiplicity' function $g:\mathcal{A}\to\mathbb{C},\ \alpha\mapsto g_\alpha$. Taking $\mathcal{A}=\{e_i-e_j\mid 1\leq i<j\leq N\}$ and $g_\alpha=g$, we recover the Schr\"odinger operator \eqref{qH}. More generally, we can follow Olshanetsky and Perelomov (1983) and allow $\mathcal{A}$ to be a positive half of an arbitrary root system while requiring $g$ to be Weyl-group invariant. In all of these cases, it is well-known that the Schr\"odinger operator $\hat{H}_{\mathcal{A}}$ is integrable when the potential function $V$ is of type $\text{I}$-$\text{IV}$. In fact, even more general examples have been found, such as the deformed CMS systems discovered by Chalykh, Feigin and Veselov (1998). These systems are associated with particular deformations of root systems and have important applications to conformal field theory, Lie superalgebras, random matrices and more. Despite all of these examples and considerable research efforts, the very natural problem of determining for precisely which pairs $(\mathcal{A},g)$ the Schr\"odinger operator $\hat{H}_{\mathcal{A}}$ is integrable remains open.

\section{Further reading}

Babelon, O., Bernard, D.~and Talon, M.~2003.
Introduction to classical integrable systems. 
Cambridge Monographs on Mathematical Physics. Cambridge: Cambridge University Press.

Calogero, F.~2001.
Classical many-body problems amenable to exact treatments.
Lecture Notes in Phys.~New Ser.~m Monogr.~66.
Berlin: Springer-Verlag.

Chalykh, O.~2008.
Algebro-geometric Schr\"odinger operators in many dimensions.
Philos.~Trans.~R.~Soc.~Lond.~Ser.~A Math.~Phys.~Eng.~Sci.~366, 947--971.

Etingof, P.~2007.
Calogero–Moser systems and representation theory.
Z\"urich Lectures in Advanced Mathematics. Z\"urich: European Mathematical Society.

Forrester, P.~J.~2010.
Log-gases and random matrices.
London Mathematical Society Monographs. Princeton: Princeton University Press.

Halln\"as, M, Noumi, M., Spiridonov, V.~P.~and Warnaar, S.~O.~(eds.). 2020.
Proceedings of the 2019 workshop Elliptic integrable systems, special functions and quantum field theory.
SIGMA 16.

Koornwinder, T.~H.~and Stokman, J.~V.~(eds.). 2021.
Encyclopedia of special functions: Multivariable special functions. Cambridge: Cambridge University Press.

Macdonald, I.~G.~1995.
Symmetric functions and Hall polynomials, 2nd edn.
New York: Oxford University Press.

Mironov, A.~and Morozov, A.~2023.
On the status of DELL systems.
arXiv: 2309.06403.

Nekrasov, N.~1999.
Infinite-dimensional algebras, many-body systems and gauge theories.
Moscow Seminar in Mathematical Physics, 263--299, 
Amer.~Math.~Soc. Transl.~Ser.~2, 191, Adv.~Math.~Sci., 43.
Providence, RI: American Mathematical Society.

Olshanetsky, M.~A.~and Perelomov, A.~M.~1983.
Quantum integrable systems related to Lie algebras.
Phys.~Rep.~94, 313--404.

Perelomov, A.~M.~1990.
Integrable systems of classical mechanics and Lie algebras.
Basel: Birkh\"auser Verlag.

R\"osler, M.~2003.
Dunkl operators: theory and applications.
In: Koelink, E.~and Van Assche, W.~(eds.) Orthogonal polynomials and special functions, Lecture Notes in Math.~1817, 93--135.
Berlin: Springer.

Ruijsenaars, S.~N.~M.~1994.
Systems of Calogero--Moser type.
In: Semenoff, G.~and Vinet, L.~(eds.) Proceedings of the 1994 Banff summer school Particles and fields, pp.~251--352. New York: Springer.

Ruijsenaars, S.~N.~M.~2004.
Elliptic integrable systems of Calogero-Moser type: A survey.
In: Noumi, M.~and Takasaki, K.~(eds.) Proceedings of the 2004 Kyoto Workshop on Elliptic integrable systems, Rokko Lectures in Math.~18, Dept.~of Math., Kobe Univ., 201--221.

\section{References}

Airault, H., McKean, H.~P.~and Moser, J.~1977.
Rational and elliptic solutions of the Korteweg--de Vries equation and a related many-body problem. 
Comm.~Pure Appl.~Math. 30, 95--148.

Belousov, N., Derkachov, S., Kharchev, S.~and Khoroshkin, S.~2023.
Baxter operators in Ruijsenaars hyperbolic system III. Orthogonality and completeness of wave functions.
arXiv:2307.16817.

Braden, H.~W., Marshakov, A., Mironov, A.~and Morozov, A.~2000.
On double-elliptic integrable systems 1. A duality argument for the case of $SU(2)$.
Nuclear Phys.~B 573, 553--572.

Buchstaber, V.~M., Felder, G.~and Veselov, A.~P.~1994.
Elliptic Dunkl operators, root systems, and functional equations. 
Duke Math.~J.~76, 885--911.

Calogero, F.~1971.
Solution of the one-dimensional N-body problems with quadratic and/or inversely quadratic pair potentials.
J.~Math.~Phys.~12, 419--436.

Chalykh, O.~2000.
Bispectrality for the quantum Ruijsenaars model and its integrable deformation.
J.~Math.~Phys.~41, 5139--5167.

Chalykh, O.~2019.
Quantum Lax Pairs via Dunkl and Cherednik Operators.
Comm.~Math.~Phys.~369, 261--316.

Chalykh, O. A., Feigin, M.~and Veselov, A. P.~(1998).
New integrable generalizations of Calogero--Moser quantum problem.
J.~Math.~Phys.~39, 695--703.

Chalykh, O. A.~and Veselov, A. P.~1990.
Commutative rings of partial differential operators and Lie algebras. 
Comm.~Math.~Phys.~126, 597--611.

Cherednik, I.~1991.
A unification of Knizhnik--Zamolodchikov equations and Dunkl operators via affine Hecke algebras.
Invent.~Math.~106, 411--432.

van Diejen, J.~F.~1994.
Integrability of difference Calogero--Moser systems.
J.~Math.~Phys.~35, 2983--3004.

Dunkl, C.~F.~1989.
Differential-difference operators associated to reflection groups. 
Trans.~Amer.~Math.~Soc.~311, 167--183.

Dunkl, C.~1992.
Hankel transforms associated to finite reflection groups. Hypergeometric functions on domains of positivity, Jack polynomials, and applications (Tampa, FL, 1991), 123--138.
Contemp.~Math.~138.
Providence, RI: American Mathematical Society.

Etingof, P., Felder, G., Ma, X.~and Veselov, A.~2011.
On elliptic Calogero--Moser systems for complex crystallographic reflection groups.
J.~Algebra 329, 107--129.

Feh\'er, L., Marshall, I.~2019.
Global description of action-angle duality for a Poisson--Lie deformation of the trigonometric $\text{BC}_n$ Sutherland system.
Ann.~Henri Poincar\'e 20, 1217--1262.

Fock, V., Gorsky, A., Nekrasov, N.~and Rubtsov, V.~2000.
Duality in integrable systems and gauge theories. 
J.~High Energy Phys.~2000,Paper 28, 40 pp.

Halln\"as, M.~and Ruijsenaars, S.~2014.
Joint eigenfunctions for the relativistic Calogero--Moser Hamiltonians of hyperbolic type: I. First steps.
Int.~Math.~Res.~Not. IMRN 2014, 4400--4456.

Heckman, G.~J.~1991.
An elementary approach to the hypergeometric shift operators of Opdam.
Invent.~Math.~103, 341--350.

Heckman, G.~J.~and Opdam, E.~M.~1987.
Root systems and hypergeometric functions. I. 
Compositio Math.~64, 329--352.

Jack, H.~1970.
A class of symmetric polynomials with a parameter.
Proc.~R.~Soc. Edinburgh Sect.~A 69 (1970/1971), 1--18.

de Jeu, M.~F.~E.~1993.
The Dunkl transform.
Invent.~Math.~113, 147--162.

Kazhdan, D., Kostant, B.~and Sternberg S.~1978.
Hamiltonian group actions and dy-namical systems of Calogero type
Comm.~Pure Appl.~Math.~31, 481--507.

Komori, Y.~and Takemura, K.~2002.
The perturbation of the quantum Calogero-Moser-Sutherland system and related results.
Comm.~Math.~Phys.~227, 93--118.

Koornwinder, T.~H.~1992.
Askey--Wilson polynomials for root systems of type $BC$.
Hypergeometric functions on domains of positivity, Jack polynomials, and applications (Tampa, FL, 1991), 189--204.
Contemp.~Math.~138.
Providence, RI: American Mathematical Society.

Koroteev, P.~and Shakirov, S.~2000.
The quantum DELL system.
Lett.~Math. Phys.~110, 969--999.

Krichever, I.~M.~1980.
Elliptic solutions of the Kadomtsev-Petviashvili equation and integrable systems of particles.
Funktsional.~Anal.~i Prilozhen.~14, 45--54.

Langmann, E.~2014.
Explicit solution of the (quantum) elliptic Calogero-- Sutherland model.
Ann.~Henri Poincar\'e 15, 755--791.

Macdonald, I.~G.~1988.
A new class of symmetric functions.
Publ.~I.R.M.A.~Strasbourg, Actes $20^{e}$ S\'eminaire Lotharingien, 131--171.

Macdonald, I.~G.~2000.
Orthogonal polynomials associated with root systems.
S\'eminaire Lotharingien de Combinatoire 45 (2000), Article B45a

Marchioro, C.~1970.
Solution of a three-body scattering problem in one dimension.
J.~Math.~Phys.~11, 2193--2196.

Moser, J.~1975.
Three integrable Hamiltonian systems connected with isospectral deformations. 
Advances in Math.~16, 197--220. 

Olshanetsky, M. A.~and Perelomov, A. M.~1977.
Completely integrable Hamiltonian systems connected with semisimple Lie algebras. 
Invent.~Math.~37 (1976), 93--108.

Opdam, E.~M.~1993.
Dunkl operators, Bessel functions and the discriminant of a finite Coxeter group. 
Compositio Math.~85, 333--373.

Polychronakos, A.P.~1992.
Exchange operator formalism for integrable systems of particles.
Phys.~Rev.~Letters, 703--705.

Ruijsenaars, S.~N.~M.~1987.
Complete integrability of relativistic Calogero--Moser systems and elliptic function identities. 
Comm.~Math.~Phys.~110, 191--213.

Ruijsenaars, S.~N.~M.~1988.
Action-angle maps and scattering theory for some finite-dimensional integrable systems. I. The pure soliton case.
Comm. Math. Phys. 115, 127--165.

Ruijsenaars, S.~N.~M.~1995.
Action-angle maps and scattering theory for some finite-dimensional integrable systems. III. Sutherland type systems and their duals.
Publ.~RIMS Kyoto Univ.~31, 247--353.

Ruijsenaars, S.~N.~M.~and Schneider, H.~1986.
A new class of integrable systems and its relation to solitons. 
Ann.~Physics 170, 370--405.

Shiraishi, J.~2019.
Affine screening operators, affine Laumon spaces and conjectures concerning non-stationary Ruijsenaars functions.
J.~Integrable Syst.~4, xyz010, 30 pp.

Sutherland, B.~1971.
Exact results for a quantum many-body problem in one dimension.
Phys.~Rev.~A 4, 2019--2021.

Toda, M.~1967.
Vibration of a chain with a nonlinear interaction.
J.~Phys.~Soc. Japan 22, 431--436.

Ujino, H., Hikami, K..~and Wadati, M.~1992.
Integrability of the quantum Calogero--Moser model.
J.~Phys.~Soc.~Japan 61, 3425--3427.

\end{document}